\documentclass[12pt]{article}
\usepackage{amsfonts}
\usepackage{amssymb}
\usepackage{color}
\usepackage{graphicx}

\newcommand{\be}{\begin{equation}}
\newcommand{\en}{\end{equation}}
\newcommand{\bea}{\begin{eqnarray}}
\newcommand{\ena}{\end{eqnarray}}
\newcommand{\beano}{\begin{eqnarray*}}
\newcommand{\enano}{\end{eqnarray*}}

\catcode `\@=11 \@addtoreset{equation}{section}
\def\theequation{\arabic{section}.\arabic{equation}}
\catcode `\@=12
\begin{document}


\begin{center}
{\Large \textbf{Towards a formalization of a two traders market with
information exchange}}

\vspace{2mm}

{\large F. Bagarello}
\vspace{1mm}\\[0pt]
DEIM, Scuola Politecnica,\\[0pt]
Universit\`{a} di Palermo, I - 90128 Palermo, and INFN, Torino, Italy\\[0pt]
E-mail: fabio.bagarello@unipa.it\\[0pt]
home page: www.unipa.it/fabio.bagarello\\[5pt]
{\large E. Haven}
\vspace{1mm}\\[0pt]
School of Management {and Institute of Finance},\\[0pt]
University of Leicester, Leicester, U.K.\\[0pt]
E-mail: e.haven@le.ac.uk \vspace{2mm}\\[0pt]
\end{center}


\begin{abstract}
This paper shows that Hamiltonians and operators can also be put to good use
even in contexts which are not purely physics based. Consider the world of
finance. The work presented here {models a two traders system with
information exchange with the help of four fundamental operators: cash and
share operators; a portfolio operator and an operator reflecting the loss of
information. An information Hamiltonian is considered and an additional
Hamiltonian is presented which reflects the dynamics of selling/buying
shares between traders. An important result of the paper is that when the
information Hamiltonian is zero, portfolio operators commute with the
Hamiltonian and this suggests that the dynamics are really due to the
information. Under the assumption that the interaction and information terms
in the Hamiltonian have similar strength, a perturbation scheme is
considered on the interaction parameter. Contrary to intuition, the paper
shows that up to a second order in the interaction parameter, a key factor
in the computation of the portfolios of traders will be the initial values
of the loss of information (rather than the initial conditions on the cash
and shares). Finally, the paper shows that a natural outcome from the
inequality of the variation of the portfolio of trader one versus the
variation of the portfolio of trader two, begs for the introduction of
`good' and `bad' information. It is shown that `good' information is related
to the reservoirs (where an infinite set of bosonic operators are used) which model rumors/news
and external facts, whilst `bad' information is associated with a set of two
modes bosonic operators. }
\end{abstract}

\vfill

\newpage


\section{Motivations}

The application of techniques from physics to areas outside of its natural
remit, such as economics and finance, is not new. In the 1960's the famous
Harvard economist Nicolas Georgescu-Roegen \cite{nicolas} considered the
concept of entropy in economics. In the 1990's the econophysics movement
started by luminaries like Eugene Stanley \cite{eugene} and Bouchaud \cite%
{bouchaud}, became an important field where techniques from statistical
mechanics were very fruitful in understanding some difficult problems in
finance and macro-economics. {An outstanding paper which argues in a very
informed way about how physics based ideas can be of benefit to
understanding a plethora of concepts in the area of complex systems is by
Kwapien and Dro\.{z}d\.{z} \cite{kd}. In that paper the authors remark that
one could guess, from a theoretical point of view that \textquotedblleft all
the laws of financial economics must be a strict mathematical consequence of
the four fundamental interactions among the elementary
particles\textquotedblright .\ But the authors do immediately caution, that
from a \textbf{practical}} {point of view such an approach is unworkable.
The authors remark that \textquotedblleft to fully explain the financial
market's behavior, one has to neglect the deeper levels of organization
without any meaningful loss of information\textquotedblright .\ Kwapien and
Dro\.{z}d\.{z} \cite{kd} ask the question, within the context of real stock
data, what part of the eigenvalue spectrum of the correlation matrix (which
contains a Wishart matrix) contains information about non-trivial
correlations. The authors find that, again within the context of stock
markets, noise and collectivity (i.e. based on a large number of non-linear
interacting constituents) are in a dynamical balance with each other, and
this typifies complex systems. The rationale for introducing some ideas out
of quantum mechanics is also highlighted in the same paper \cite{kd}, where
the authors do mention what the interpretation might be of a price of an
asset between two consecutive transactions. Indeed the non-observed price
could be associated with a quantum mechanical measurement problem. The
authors cite the work of Schaden \cite{ms} in that regard. Other works
appeared also in that area. In the field of game theory, the solution space
of even very elementary games can be enriched when a quantum mechanical
interpretation is considered. In the paper by Piotrowski and S\l adkowski
\cite{psi} the authors invite the reader to consider what happens when
trading strategies are allowed to be entangled. The paper proposes the idea
of a quantum strategy. An important paper, again by these authors \cite{psg}%
, investigates} a crucial aspect for improving our understanding of
financial markets: Information. {The concept of information is well
formalized in physics and the paper shows that the formalization of
information via the metric structures can be a very good step in the right
direction for a deeper theoretical understanding of financial markets. In
effect we will below, in the motivation for this paper, address the
information issue a little more.}

{The }area of research which applies techniques from quantum mechanics to a
variety of problems in the social sciences, {can actually be }traced back to
the 1950's, with the discussions physics Nobelist Pauli had with the well
known psychologist Jung, on basic issues such as how complementarity in
quantum physics can have `some' existence in psychology \cite{meier}, \cite%
{ns}. The level of effectiveness by which quantum mechanical techniques have
been able to shed further light on thorny problems in a variety of areas in
the social sciences, varies somewhat. In psychology, there is sizable
research-momentum in the particular field which actively uses probability
interference to decision making paradoxes in economics and psychology \cite%
{busemeyer}, \cite{ehti}. In the area of information retrieval, research
advances are also made \cite{cohen}. In finance, which is the area of
application of this current paper, progress has been made on importing the
quantum physical machinery in an attempt to augment the modelling of
information. Other work in this area has also looked at how potential
functions (within the quantum mechanical setting) can adopt financial
meaning \cite{zhang}, \cite{belal}.

In a recent paper, \cite{baghav}, the authors have considered the way in
which information reaching different traders of a (simplified) stock market
influences the behavior of the traders, \textbf{before they begin to trade}.
In other words, we have considered what happens before the market opens, and
in which way the strategy of the traders is generated. In this description
we have used tools which are originally encountered in the microscopic
world, and which have been proven to be useful also in the description of
different classical systems, see \cite{bagbook} for a recent review. In
particular, a special role is played by an operator, the Hamiltonian of the
system, which is used to deduce the dynamics of those quantities we are
interested in, the so-called \emph{observables} of the model.

In some older papers of one of us (F.B.), \cite{bag1}-\cite{bag4}, the role
of information was, in a certain sense, simply incorporated by properly
choosing some of the constants defining the Hamiltonian of the system we
were considering. The Hamiltonian is adopted to mimic and describe the
interactions between the traders, \cite{bagbook}. On the other hand, E.H.
and his coworkers, {following the original idea of \cite{khren1}, }%
considered the role of information for stock markets, \cite{hav1}-\cite{hav2}%
, mainly adopting the Bohm view of quantum mechanics, where the information
is carried by a pilot wave function $\Psi (x,t)$, satisfying a certain Schr%
\"{o}dinger equation of motion, and which, with simple computations,
produces what in the literature is called \emph{a mental force}. This force
has to be added to the other \emph{hard} forces acting on the system,
producing a full Newton-like classical differential equation.

In \cite{baghav} we have tried to produce an unifying point of view, using
Bohmian quantum mechanics to construct a Hamiltonian $H$ in which the
information is not merely described by some parameters of $H$, but becomes
one of the dynamical variables of the system. However, in that preliminary
work, we have only considered how the information contributes to generate,
out of two \emph{equivalent} traders $\tau _{1}$ and $\tau _{2}$, two
traders which are no longer equivalent: i.e. they have used the information
to improve, as much as they can, their \emph{financial status} (the
portfolio, see below). For this reason, no interaction between $\tau _{1}$
and $\tau _{2}$ was considered in \cite{baghav}. Here we continue our
analysis adding also a possible interaction to the system. In other words,
we will see what happens in a market made of $\tau _{1}$ and $\tau _{2}$,
when they interact and are also subjected to a flux of information coming
from the market itself and from the outer world. As one can expect, this is
quite a hard problem to be discussed in its full generality, and in fact we
will consider, along the way, some useful assumptions which will allow us to
deduce an approximate analytical solution for the problem.

{To clarify our main ideas, we propose here a list of six succinct points
which are those motivating our present analysis. We keep specifically in
mind the `un-convinced' or `sceptic' reader. }

\begin{itemize}
\item {First, the Hamiltonians which are used in the paper are introducing
dynamics in the model in a \textquotedblleft natural\textquotedblright\ way.
We can explicitly claim that the Hamiltonians considered here are receiving
an economics based interpretation. Important work in the literature has also
referred to the use of a Hamiltonian framework in a social science
framework. In Kwapien and Dro\.{z}d\.{z} \cite{kd} reference is made to a so
called market factor which is a force acting on all stocks. As the authors
explain, this approach refers to a many-body problem which can lead to the
use of a Hamiltonian. In their paper Piotrowski and S\l adkowski \cite{psq}
use a Hamiltonian which contains what they define as a `risk inclination
operator'. Our paper expands the Hamiltonian (relative to our first paper
(Bagarello and Haven \cite{baghav})) to a Hamiltonian which now also models
interaction, even if in a very simplified form. Whilst our first paper had
an absence of interaction between traders, and this current paper explicitly
allows for interaction between traders, it should be stressed that \textbf{%
even in the absence of interaction} there was quite some richness in the
first paper. The limit on number of traders was of course irrelevant given
the absence of interaction, but even with this absence, we were very
concerned to discuss what happens before trading begins and after the rumors
have reached traders. Our first paper also actively studied the situation of
two traders who are no longer completely equivalent. In this paper, we think
that it is quite important to observe that we can now divide information,
using the expanded Hamiltonian framework, into two sets of information - bad
and good information. Information is seen as a dynamical variable and it
thus has a role in the Hamiltonian itself. This leads us to make a plea
about how useful in fact quantum mechanical concepts in social science can
be. We believe that the modelling of information is a very big advantage
that the quantum formalism has to offer when considering applications
outside of the remit of quantum mechanics. We want to hint to the use of
Fisher information (well known in economics via the so called Cramer-Rao
bound) and the intimate relationship which exists between the minimization
of Fisher information and the Schr\"{o}dinger equation (see Hawkins and
Frieden \cite{HF}). Please see also point five below. We also can mention
the relationship which has been argued for between Fisher information and a
specific type of potential (see Reginatto \cite{REG}, Haven and Khrennikov
\cite{HAV4}).}

\item {Second, the use of non-commuting operators has been investigated in
the finance environment. In Segal and Segal \cite{SEGAL} it is shown that
such operators should be used to describe the time dependence of the price
of shares and its forward time derivative. {The motivation is purely
economical: if one trader knows exactly both these quantities, he could earn
a virtually enormous amount of money. Since this does not happen, it is
reasonable to replace functions of time with time-depending, non commuting,
operators.}}

\item {Third, bosonic operators have a financial meaning in this paper and
the reason why such bosonic operators are coming in a natural way in our
financial set up is linked to the fact that the operator can assume a very
large set of discrete values. {This gives us the possibility to describe, in
a rather natural way, the portfolios (see below) of the traders.}}

\item {Fourth, the reservoir with which the traders interact produces a
system with infinite degrees of freedom. }

\item {Fifth, we can, as we have expressed in several footnotes in the
current paper, look at the measure of loss of information within the context
of a traded financial payoff function and Fisher information (which we
mentioned in our first point above).}

\item {Sixth, we probably should also mention that very strong connections
have been established between the Schr\"{o}dinger and the Black-Scholes
equations, \cite{baa}. This is surely another indication of the relevance of
quantum mechanics in economics.}
\end{itemize}

{In summary, we think both quantum mechanics and financial markets benefit
from our approach. From a quantum mechanical point of view, we show that
uses can be made of elementary concepts outside of the natural remit of
quantum mechanics. We believe that the above 6 points provide for good
arguments why this current study can provide benefits for better
understanding financial markets. Very few models in economics will use
Hamiltonians which have an information and interaction component to describe
dynamics. We can only make `baby-steps' at this point in time, but it should
be seen as a credible argument, that given the extremely powerful machinery
quantum mechanics really is, it may not be impossible to harness that power
also within a social science domain. It is surely not the case, that finance
and economics should not be receptive to new models. Quite the contrary, for
its own survival, it should be open to models coming from other areas of
inquiry. Many models in the finance literature are often extremely simple
too. Often the assumptions underlying those models make the applicability of
the model to be very constrained. We have been very up-front in this paper
with our assumptions. }

The paper is organized as follows: in the next section we propose our model
and we discuss some of its most important aspects. In particular, we deduce
the relevant equations of motion. In Section III we propose a perturbative
approach to deduce the approximate solution of these equations. Section IV
contains our conclusions.

\section{The model}

The model we are interested here extends the one originally proposed in \cite%
{baghav}, adding an explicit interaction term between the traders. We begin
by defining the following Hamiltonian, already considered in \cite{baghav}:

\begin{equation}
\left\{
\begin{array}{ll}
\mathfrak{H}=H_{0}+H_{inf}, &  \\
H_{0}=\sum_{j=1}^{2}\left[ \omega _{j}^{s}\hat{S}_{j}+\omega _{j}^{c}\hat{K}%
_{j}+\Omega _{j}\hat{I}_{j}+\int_{\mathbb{R}}\Omega _{j}^{(r)}(k)\hat{R}%
_{j}(k)\,dk\right] , &  \\
H_{inf}=\sum_{j=1}^{2}\left[ \lambda _{inf}\left( i_{j}(s_{j}^{\dagger
}+c_{j}^{\dagger })+i_{j}^{\dagger }(s_{j}+c_{j})\right) +\gamma _{j}\int_{%
\mathbb{R}}(i_{j}^{\dagger }r_{j}(k)+i_{j}r_{j}^{\dagger }(k))\,dk\right] ,%
\label{21} &
\end{array}%
\right.
\end{equation}%
where $\hat{R}_{j}(k)=r_{j}^{\dagger }(k)r_{j}(k)$, $\hat{S}%
_{j}=s_{j}^{\dagger }s_{j}$, $\hat{K}_{j}=c_{j}^{\dagger }c_{j}$ and $\hat{I}%
_{j}=i_{j}^{\dagger }i_{j}$, and the following canonical commutation
relations (CCR's) are assumed,
\begin{equation}
\lbrack s_{j},s_{l}^{\dagger }]=[c_{j},c_{l}^{\dagger
}]=[i_{j},i_{l}^{\dagger }]=1\!\!1\delta _{j,l},\quad \lbrack
r_{j}(k),r_{l}^{\dagger }(q)]=1\!\!1\delta _{j,l}\delta (k-q),  \label{21bis}
\end{equation}%
all the other commutators being zero. Moreover $\omega _{j}^{s}$, $\omega
_{j}^{c}$, $\Omega _{j}$, $\lambda _{inf}$ and $\gamma _{j}$ are real
constants, while $\Omega _{j}^{(r)}(k)$, $j=1,2$, are two real-valued
functions. Each bosonic operator has a different meaning in the present
context, which is explained in detail in \cite{baghav}: $c_{j}$, $%
c_{j}^{\dagger }$ and $\hat{K}_{j}$ are \emph{cash operators}. They
respectively lower, increase and count the units of cash in the portfolio of
$\tau _{j}$, see below. Analogously $s_{j}$, $s_{j}^{\dagger }$ and $\hat{S}%
_{j}$ are \emph{share operators}. They lower, increase and count the number
of shares in the portfolio of $\tau _{j}$. Incidentally, we notice that, to
make the notation simple, we are assuming that our market consists of a
single type of shares. This is not a major constraint, and it could be
avoided. However, we will not do it here. The operator $i_{j}^{\dagger }$
increases the \emph{lack of information} (LoI) of $\tau _{j}$, while $i_{j}$
decreases it. Of course, the higher the value of the eigenvalues of the
number-like operator $\hat{I}_{j}$, the less $\tau _{j}$ knows about what is
going on in the market. In other words, to be more efficient, the trader
should have a low LoI, i.e. he should be somehow associated to a small
eigenvalue of $\hat{I}_{j}$. In our model we also have a reservoir, which
models the set of all the rumors, news, and external facts which, all
together, concretely create the final information and, therefore, fix the
values of the LoI's of the two traders. The reservoir\footnote{%
In financial economics, a distinction is often made between so called
`private information' and `public information'. The reservoir here contains
public and private information. Distinguishing those types of information
can be fruitful as they do implicitly call up notions such as `financial
efficiency' where the strongest form of efficiency would say that all prices
contain both public and private information. This is thus the point of view
taken in this paper.} is described here by the bosonic operators $r_{j}(k)$,
$r_{j}^{\dagger }(k)$ and $\hat{R}_{j}(k)$, which depend on a real variable,
$k\in \mathbb{R}$.

The Hamiltonian $\mathfrak{H}$ contains a free \emph{canonical} part $H_{0}$%
. By this we mean that $H_0$ is the typical quadratic Hamiltonian used in
quantum many-body systems, when they are described in second quantization.
The main characteristic of $H_0$ is that, whenever our system is described
only by $H_0$, i.e. when we put $H_{inf}=0$, all the number operators ($\hat
S_j$, $\hat K_j$ and so on) stay constant in time: so, from the point of
view of our \emph{observables}, \cite{bagbook}, the market looks static.
However, this is not really so, since non-observable operators may still
evolve in time.

For what concerns $H_{inf}$, let us now consider separately its two
contributions. They respectively describe the following: when the LoI
increases, the value of the portfolio decreases (because of $i_{j}^{\dagger
}(s_{j}+c_{j})$) and vice-versa (because of $i_{j}(s_{j}^{\dagger
}+c_{j}^{\dagger })$)\footnote{%
This first contribution to $H_{inf}$ can also be obtained via a slightly
different route (\cite{hav3}) where a quantum mechanical (like) wave
function is seen as carrier of information and upon it travelling towards a
potential function (the payoff function) it may decay or not depending on
the position of the potential versus total energy. Total energy is
considered as capturing public information, whilst the wave function carries
private information. If we assume the portfolio to be a payoff function
which maps a domain of prices of shares onto a level of profit then one can
model incoming information, before a profit position is taken, as decaying
in a way which will depend on the level of the profit. We show that if we
restrict this domain of prices to be very narrow, then the higher the level
of profit of the payoff function the lower the LoI and the lower the level
of profit, the higher the LoI. The change in LoI could be measured via the
comparison of two Fisher information measures.}. Moreover, the LoI increases
when the "value" of the reservoir decreases (this is the meaning of $%
i_{j}^{\dagger }r_{j}(k)$), and, viceversa, decreases when the "value" of
the reservoir increases\footnote{%
If we consider again \cite{hav3}, we can obtain a similar result - but again
in a different setting. Consider the reservoir to be total energy and let
there be a payoff function (which is a potential function) with a large
domain of prices of shares. Set first the level of total energy vis a vis
the payoff function such that the incoming quantum (like) mechanical wave
function does not decay and calculate the Fisher information. Now reduce the
level of total energy such that the incoming quantum (like) mechanical wave
function will decay and measure the Fisher information. If the domain of
prices is sufficiently large, once can show indeed that the LoI increases
when the value of the reservoir decreases.}. Considering, for example, the
contribution $i_{j}r_{j}^{\dagger }(k)$ in $H_{inf}$, we see that the LoI
decreases (so that the trader is \emph{better informed}) when a larger
amount of news, rumors, etc. reaches the trader. Notice that, in $\mathfrak{H%
}$, no interaction between $\tau _{1}$ and $\tau _{2}$ is considered, yet.
As in \cite{baghav}, to produce a reasonably simple model, we will assume
that the price of the share is constant in time, and we fix this constant to
be one. Of course, this is a strong limitation of the model, but it is
useful to allow to get some analytical expression for the time evolution of
the portfolios of the traders. {Other possibilities exist, but, not
surprisingly, produce more complicated models: one could consider the price
of the share as a dynamical variable of the system. This is what we really
would like to do, but it is very hard to implement this possibility in a
realistic way. A simpler possibility is to consider the price as an external
field, deduced out of experimental data. Both these possibilities are
discussed in \cite{bagbook}.} We will come back on this aspect of the model
later on.

In \cite{baghav} $\mathfrak{H}$ was exactly the objective of our interest,
since we were not considering the interaction between the traders. Here, on
the other hand, this is exactly one of the aspects which is interesting for
us. For this reason, our full model is described by the following
Hamiltonian:
\begin{equation}
\left\{
\begin{array}{ll}
H=\mathfrak{H}+H_{int}, &  \\
H_{int}=\lambda\left(s_1c_1^\dagger s_2^\dagger c_2+s_1^\dagger
c_1s_2c_2^\dagger\right). \label{22} &
\end{array}%
\right.
\end{equation}
The meaning of $H_{int}$ is the following: $s_1c_1^\dagger s_2^\dagger c_2$
describes the fact that $\tau_1$ is selling a share to $\tau_2$. For this
reason, the number of the shares in his portfolio decreases of one unit (and
this is the meaning of $s_1$) while his cash increases of one unit (because
of $c_1^\dagger$), since the price of the share is assumed here to be one%
\footnote{%
In some older models, \cite{bagbook}, $c_1^\dagger$ was replaced by ${%
c_1^\dagger}^{\hat P}$, where $\hat P$ is the price operator.}. After the
interaction, $\tau_2$ has one more share ($s_2^\dagger$), but one less unit
of cash ($c_2$). Of course, $H_{int}$ also contains the adjoint
contribution, which describes the opposite situation: $\tau_2$ sells a share
to $\tau_1$. Therefore, for obvious reasons, $s_jc_j^\dagger$ and $%
s_j^\dagger c_j$ can be collectively called \emph{the selling and buying
operators}, respectively. $\lambda$ is an interaction parameter: if $%
\lambda=0$, $\tau_1$ and $\tau_2$ do not interact, and we go back to our
analysis in \cite{baghav}.

It is worth stressing that our choice of Hamiltonian is not compatible with
the fact that the amount of cash and the number of shares are preserved
during the time evolution. This is a simple consequence of the fact that,
calling $\hat{K}=\sum_{j=1}^{2}\hat{K}_{j}$ and $\hat{S}=\sum_{j=1}^{2}\hat{S%
}_{j}$ the total cash and number of shares operators of the market, they do
not commute with $H$. In fact, in particular, they do not commute with $%
H_{inf}$: $[H,\hat{K}]\neq 0$, $[H,\hat{S}]\neq 0$. Hence, we are allowing
here for bankruptcy. Moreover, we are not assuming that the cash is only
used to buy shares, so that it needs not to be preserved in time. However,
some other self-adjoint operators are preserved during the time evolution.
These operators are $\hat{M}_{j}=\hat{S}_{j}+\hat{K}_{j}+\hat{I}_{j}+\hat{R}%
_{j}=\hat{\Pi}_{j}+\hat{I}_{j}+\hat{R}_{j}$, $j=1,2$, where $\hat{R}%
_{j}=\int_{\mathbb{R}}r_{j}^{\dagger }(k)r_{j}(k)\,dk$ and $\hat{\Pi}_{j}=%
\hat{S}_{j}+\hat{K}_{j}$ is what we call the \emph{portfolio operator} of $%
\tau _{j}$, which is simply the sum of the trader $j$'s amount of cash and
number of shares (and, being the price of each share equal to one, also
their value). Then we can check that $[H,\hat{M}_{j}]=0$, $j=1,2$. This
implies that what is constant in time is the sum of the portfolio, the LoI
and the \emph{reservoir input} of each single trader\footnote{%
The existence of conserved quantities has proven to be useful, among other
reasons, also to check that the numerical schemes adopted to solve the
equations of the system work properly, \cite{ff}.}.

What we are willing to deduce is the time evolution of the portfolio
operators: $\hat\Pi_j(t)=\hat K_j(t)+\hat S_j(t)$, $j=1,2$. As we have
already noticed, because of our working assumption about the price of the
shares, the mean value of $\hat\Pi_j(t)$ represents for us the \emph{richness%
} of $\tau_j $. The mean value, as widely discussed in \cite{bagbook}, has
to be taken with respect to vectors which are eigenstates of (all) the
number operators of the system, with eigenvalues corresponding to the
initial conditions of the system. We will see explicitly how this
computation works later on.

Once we have the Hamiltonian, we can deduce the differential equations we
are interested in by adopting the standard quantum mechanical Heisenberg
approach: $\dot{X}=i[H,X]$. In this way, we get the following set of
equations:
\begin{equation}
\left\{
\begin{array}{ll}
\frac{d}{dt}s_{k}(t)=-i\omega _{k}^{s}s_{k}(t)-i\lambda
_{inf}\,i_{k}(t)-i\lambda c_{k}(t)s_{\overline{k}}(t)c_{\overline{k}%
}^{\dagger }(t), &  \\
\frac{d}{dt}c_{k}(t)=-i\omega _{k}^{c}c_{k}(t)-i\lambda
_{inf}\,i_{k}(t)-i\lambda s_{k}(t)c_{\overline{k}}(t)s_{\overline{k}%
}^{\dagger }(t), &  \\
\frac{d}{dt}i_{k}(t)=-i\Omega _{k}i_{k}(t)-i\lambda
_{inf}(s_{k}(t)+c_{k}(t))-i\gamma _{k}\int_{\mathbb{R}}r_{k}(q,t)\,dq &  \\
\frac{d}{dt}r_{k}(q,t)=-i\Omega _{k}^{(r)}(q)\,r_{k}(q,t)-i\gamma
_{k}\,i_{k}(t),\label{23} &
\end{array}%
\right.
\end{equation}%
where the \emph{denial of $k$}, $\overline{k}$, is seen as follows: if $k=1$%
, then $\overline{k}=2$. On the other hand, when $k=2$, then $\overline{k}=1$%
. With respect to the equations deduced in \cite{baghav}, in this paper we
deduce two highly nonlinear contributions in the first two equations above.
Not surprisingly, we are not able to solve the system exactly. Still, we
will produce a perturbative solution which, we believe, is of some interest.

Our first step consists in rewriting the last equation in its integral form:
\[
r_{k}(q,t)=r_{k}(q)e^{-i\Omega _{k}^{(r)}(q)t}-i\gamma
_{k}\int_{0}^{t}i_{k}(t_{1})e^{-i\Omega _{k}^{(r)}(q)(t-t_{1})}\,dt_{1},
\]%
and replacing this in the differential equation for $i_{k}(t)$. Assuming
first that $\Omega _{k}^{(r)}(q)$ is linear in $q$, $\Omega
_{k}^{(r)}(q)=\Omega _{k}^{(r)}\,q$, \cite{bagbook}, we deduce that
\begin{equation}
\frac{d}{dt}i_{k}(t)=-\left( i\Omega _{k}+\frac{\pi \gamma _{k}^{2}}{\Omega
_{k}^{(r)}}\right) i_{k}(t)-i\gamma _{k}\int_{\mathbb{R}}r_{k}(k)e^{-i\Omega
_{k}^{(r)}\,q\,t}\,dq-i\lambda _{inf}(s_{k}(t)+c_{k}(t)).  \label{24}
\end{equation}%
So far, our computations are exact. However, to find some analytical
solution, we are forced to consider some approximations and to perform some
perturbative expansion. For this reason, as in \cite{baghav}, we will now
work under the assumption that the last contribution in this equation can be
neglected, when compared to the other ones. In other words, we are taking $%
\lambda _{inf}$ to be very small. However, our procedure is much better than
simply considering $\lambda _{inf}=0$ in $H$ above, since we will keep
memory of its effects in the first two equations in (\ref{23}). Solving now (%
\ref{24}) in its simplified expression, we get
\begin{equation}
i_{k}(t)=e^{-\left( i\Omega _{k}+\frac{\pi \gamma _{k}^{2}}{\Omega _{k}^{(r)}%
}\right) t}\left( i_{k}(0)-i\gamma _{k}\int_{\mathbb{R}}r_{k}(q)\rho
_{k}(q,t)\,dq\right) ,  \label{2a}
\end{equation}%
where
\[
\rho _{k}(q,t)=\int_{0}^{t}e^{\left[ i(\Omega _{k}-\Omega _{k}^{(r)}q)+\frac{%
\pi \gamma _{k}^{2}}{\Omega _{k}^{(r)}}\right] t_{1}}\,dt_{1}=\frac{e^{\left[
i(\Omega _{k}-\Omega _{k}^{(r)}q)+\frac{\pi \gamma _{k}^{2}}{\Omega
_{k}^{(r)}}\right] t}-1}{i(\Omega _{k}-\Omega _{k}^{(r)}q)+\frac{\pi \gamma
_{k}^{2}}{\Omega _{k}^{(r)}}}.
\]%
The differential equations for $s_{j}(t)$ and $c_{j}(t)$ look now as
follows:
\begin{equation}
\left\{
\begin{array}{ll}
\dot{s}_{1}(t)=-i\omega _{1}^{s}s_{1}(t)-i\lambda
c_{1}(t)s_{2}(t)c_{2}^{\dagger }(t)-i\lambda _{inf}\,i_{1}(t), &  \\
\dot{s}_{2}(t)=-i\omega _{2}^{s}s_{2}(t)-i\lambda
c_{2}(t)s_{1}(t)c_{1}^{\dagger }(t)-i\lambda _{inf}\,i_{2}(t), &  \\
\dot{c}_{1}(t)=-i\omega _{1}^{c}c_{1}(t)-i\lambda
s_{1}(t)c_{2}(t)s_{2}^{\dagger }(t)-i\lambda _{inf}\,i_{1}(t), &  \\
\dot{c}_{2}(t)=-i\omega _{2}^{c}c_{2}(t)-i\lambda
s_{2}(t)c_{1}(t)s_{1}^{\dagger }(t)-i\lambda _{inf}\,i_{2}(t). &
\end{array}%
\right.   \label{25}
\end{equation}%
\vspace{2mm}

\textbf{Remark: }So far, the order of the various operators appearing in the
right-hand side of these equations is not important since they all commute
between them at equal time: $[c_{1}(t),s_{1}^{\dagger }(t)]=0$, for all $%
t\in \mathbb{R}$, and so on. This is a consequence of the analogous
commutation rule at $t=0$, and of the fact that the time evolution is
unitarily implemented by $H$: $X(t)=e^{iHt}X(0)e^{-iHt}$, for each dynamical
variable $X$.

\subsection{What if we remove the information?}

We devote this short subsection to briefly discuss how crucial the
information really is in our model. First we check what happens if $%
H_{inf}=0 $ in the definition of $H$. A simple computation shows that, in
this case, the differential equations deduced by this new hamiltonian
coincide exactly with those in (\ref{25}), with $i_1(t)=i_2(t)=0$. This
shows that the presence of $H_{inf}$ in $H$ is to produce something like an
external force driving the time evolution of the dynamical variables we are
interested in, $s_j(t)$ and $c_j(t)$, and $\hat\Pi_j(t)$ as a consequence.
It should be stressed that, in (\ref{25}), $i_1(t)$ and $i_2(t)$ are now
known operator-valued functions of time given in (\ref{2a}). In other words,
removing $H_{inf}$ is like removing these known forces.

Let us now look for the dynamical behavior of the two portfolio operators in
this case: in principle, we should solve the Heisenberg differential
equations in (\ref{25}) putting $\lambda _{inf}=0$. Needless to say, this
system is not trivial, and a solution could be found when one considers a
perturbation scheme for when $\lambda $ is a small parameter. However, due
to the canonical commutation rules we have assumed here, (\ref{21bis}), it
is easy to check that the dynamics of the two portfolios is trivial. In
fact, since $H=H_{0}+H_{int}$, it is a simple exercise to check that $[H,%
\hat{\Pi}_{j}]=0$, $j=1,2$. Hence, $\hat{\Pi}_{j}(t)=\Pi _{j}(0)$ for all $%
t\in \mathbb{R}$: even if the cash and the shares of the two traders may
change in time, their portfolios do not. This result seems reasonable since
our traders, receiving no information from outside the market, have no real
reason to change their original status, even if they could, in principle,
interact. However, this conclusion is strongly related to the fact that, in
our model, the price of the share stays constant in time. In fact, if this
is not so, then the portfolio of, say, $\tau _{1}$, should be defined more
reasonably as $\Pi _{1}(t):=\hat{K}_{1}(t)+\hat{P}(t)\hat{S}_{1}(t)$, $\hat{P%
}(t)$ being the value of the share at time $t$, and this operator needs not
to commute with $H$, even when $H_{inf}=0$.

\vspace{3mm}

The conclusion of this simple analysis is therefore that, in order not to
trivialize the model, $H_{inf}$ cannot be taken to be zero, so that the
equations to be solved are exactly those in (\ref{25}), but with all their
ingredients inside!

\section{The perturbative solution of the equations}

Our previous results suggest to check, first of all, that when $H_{inf}\neq0$%
, the portfolio operators do not commute with $H$. In fact, as shown before,
if they commute, there is no reason to try to solve the differential
equations, and the model is (essentially) trivial, and surely not very
interesting for us. However, luckily enough, this is not so:
\[
[H,\hat\Pi_j]=\lambda_{inf}\left(i_j^\dagger(s_j+c_j)-i_j(s_j^\dagger+c_j^%
\dagger)\right),
\]
$j=1,2$. This, again, is a measure of the relevance of the information in
our model: it is exactly the presence of $H_{inf}$ which makes the model not
trivial, not really the interaction between $\tau_1$ and $\tau_2$.

We are now ready to set up our perturbation scheme. For that, it is
convenient to define new variables $\sigma _{j}(t):=s_{j}(t)e^{i\omega
_{j}^{s}t}$ and $\theta _{j}(t):=c_{j}(t)e^{i\omega _{j}^{c}t}$, $j=1,2$. To
simplify the treatment a little bit, we also assume that $\lambda =\lambda
_{inf}$. From an economical point of view, this simply means that we are
assuming that the interaction and the information terms in $H$ have a
similar strength. Then equations (\ref{25}) become
\begin{equation}
\left\{
\begin{array}{ll}
\dot{\sigma}_{1}(t)=-i\lambda \left( \sigma _{2}(t)\theta _{1}(t)\theta
_{2}^{\dagger }(t)e^{i\hat{\omega}t}+i_{1}(t)e^{i\omega _{1}^{s}t}\right) ,
&  \\
\dot{\sigma}_{2}(t)=-i\lambda \left( \sigma _{1}(t)\theta _{1}^{\dagger
}(t)\theta _{2}(t)e^{-i\hat{\omega}t}+i_{2}(t)e^{i\omega _{2}^{s}t}\right) ,
&  \\
\dot{\theta}_{1}(t)=-i\lambda \left( \sigma _{1}(t)\sigma _{2}^{\dagger
}(t)\theta _{2}(t)e^{-i\hat{\omega}t}+i_{1}(t)e^{i\omega _{1}^{c}t}\right) ,
&  \\
\dot{\theta}_{2}(t)=-i\lambda \left( \sigma _{1}^{\dagger }(t)\sigma
_{2}(t)\theta _{1}(t)e^{i\hat{\omega}t}+i_{2}(t)e^{i\omega _{2}^{c}t}\right)
, &
\end{array}%
\right.   \label{31}
\end{equation}%
where $\hat{\omega}=\omega _{1}^{s}-\omega _{2}^{s}-\omega _{1}^{c}+\omega
_{2}^{s}$. The zero-th approximation in $\lambda $ is quite simple: $\dot{%
\sigma}_{j}^{(0)}(t)=\dot{\theta}_{j}^{(0)}(t)=0$, for $j=1,2$. Therefore,
with obvious notation, $\sigma _{j}^{(0)}(t)=\sigma _{j}^{(0)}(0)=s_{j}$ and
$\theta _{j}^{(0)}(t)=\theta _{j}^{(0)}(0)=c_{j}$, $j=1,2$, which we insert
in the right-hand side of system (\ref{31}) to deduce the first order
approximation for $\sigma _{j}(t)$ and $\theta _{j}(t)$. By introducing the
new (known) operators
\[
I_{j}^{s}(t):=\int_{0}^{t}i_{j}(t_{1})e^{i\omega _{j}^{s}t_{1}}dt_{1},\qquad
I_{j}^{c}(t):=\int_{0}^{t}i_{j}(t_{1})e^{i\omega _{j}^{c}t_{1}}dt_{1},
\]%
and by assuming that $\sigma _{j}^{(1)}(0)=s_{j}$, $\theta
_{j}^{(1)}(0)=c_{j}$ and that $\hat{\omega}\neq 0$, we get
\begin{equation}
\left\{
\begin{array}{ll}
\sigma _{1}^{(1)}(t)=s_{1}-\frac{\lambda }{\hat{\omega}}\left( e^{i\hat{%
\omega}t}-1\right) s_{2}c_{1}c_{2}^{\dagger }-i\lambda I_{1}^{s}(t), &  \\
\sigma _{2}^{(1)}(t)=s_{2}+\frac{\lambda }{\hat{\omega}}\left( e^{-i\hat{%
\omega}t}-1\right) s_{1}c_{1}^{\dagger }c_{2}-i\lambda I_{2}^{s}(t), &  \\
\theta _{1}^{(1)}(t)=c_{1}+\frac{\lambda }{\hat{\omega}}\left( e^{-i\hat{%
\omega}t}-1\right) s_{1}s_{2}^{\dagger }c_{2}-i\lambda I_{1}^{c}(t), &  \\
\theta _{2}^{(1)}(t)=c_{2}-\frac{\lambda }{\hat{\omega}}\left( e^{i\hat{%
\omega}t}-1\right) s_{1}^{\dagger }s_{2}c_{1}-i\lambda I_{2}^{c}(t). &
\end{array}%
\right.   \label{32}
\end{equation}%
It is not hard to check that this first order in our perturbation expansion
is not enough: in fact, \cite{bagbook}, in order to deduce, the (classical)
function $n_{j}(t)$, we have to compute the following mean value:
\[
n_{j}(t):=\left\langle \varphi _{\mathcal{G}_{0}},s_{j}^{\dagger
}(t)s_{j}(t)\varphi _{\mathcal{G}_{0}}\right\rangle =\left\langle \varphi _{%
\mathcal{G}_{0}},\sigma _{j}^{\dagger }(t)\sigma _{j}(t)\varphi _{\mathcal{G}%
_{0}}\right\rangle \simeq
\]%
\[
\simeq \left\langle \varphi _{\mathcal{G}_{0}},(\sigma
_{j}^{(1)}(t))^{\dagger }\sigma _{j}^{(1)}(t)\varphi _{\mathcal{G}%
_{0}}\right\rangle .
\]%
Analogously,
\[
k_{j}(t):=\left\langle \varphi _{\mathcal{G}_{0}},c_{j}^{\dagger
}(t)c_{j}(t)\varphi _{\mathcal{G}_{0}}\right\rangle \simeq \left\langle
\varphi _{\mathcal{G}_{0}},(\theta _{j}^{(1)}(t))^{\dagger }\theta
_{j}^{(1)}(t)\varphi _{\mathcal{G}_{0}}\right\rangle .
\]%
Here the vector $\varphi _{\mathcal{G}_{0}}$ is
\[
\varphi _{\mathcal{G}_{0}}=\frac{1}{\sqrt{%
n_{1}!n_{2}!k_{1}!k_{2}!I_{1}!I_{2}!}}(s_{1}^{\dagger
})^{n_{1}}(s_{2}^{\dagger })^{n_{2}}(c_{1}^{\dagger
})^{k_{1}}(c_{2}^{\dagger })^{k_{2}}(i_{1}^{\dagger
})^{I_{1}}(i_{2}^{\dagger })^{I_{2}}\varphi _{\underline{0}},
\]%
and $\varphi _{\underline{0}}$ is the vacuum of $s_{j}$, $c_{j}$ and $i_{j}$%
: $s_{j}\varphi _{\underline{0}}=c_{j}\varphi _{\underline{0}}=i_{j}\varphi
_{\underline{0}}=0$, $j=1,2$, see \cite{bagbook}. The explicit choice of the
numbers $n_{1}$, $n_{2}$, $k_{1}$, $k_{2}$ $I_{1}$ and $I_{2}$ depends on
the original (i.e., at $t=0$) status of the two traders: for example, $n_{1}$
is the number of share that $\tau _{1}$ has at $t=0$, $k_{1}$ are the units
of cash in his portfolio, at this same time, while $I_{1}$ is his LoI. Easy
computations show that, at this order in $\lambda $, $n_{j}(t)=n_{j}(0)=n_{j}
$ and $k_{j}(t)=k_{j}(0)=k_{j}$, so that each portfolio stays constant in
time: $\Pi _{j}(t)=\Pi _{j}(0)$. The conclusion is therefore that, if we
want to get some non trivial dynamics, we need to go, at least, at the
second order in the perturbation expansion.

This second order has to be deduced in the same way: we replace the first
order solution in the right-hand side of system (\ref{31}), and then we
simply integrate on time, requiring that $\sigma _{j}^{(2)}(0)=s_{j}$ and $%
\theta _{j}^{(2)}(0)=c_{j}$. Incidentally, we should observe that because of
this approximation, we get problems of ordering of the operators. In fact,
while as we have already discussed, $\sigma _{2}(t)\theta _{1}(t)\theta
_{2}^{\dagger }(t)=\theta _{1}(t)\sigma _{2}(t)\theta _{2}^{\dagger
}(t)=\sigma _{2}(t)\theta _{1}(t)\theta _{2}^{\dagger }(t)$, these
equalities are false when we replace the operators with their first, or
second, order approximations. For this reason we adopt here the following
\emph{normal ordering rule}: every time we have products of operators, we
order them considering first $s_{1}$ or $s_{1}^{\dagger }$, then $s_{2}$ or $%
s_{2}^{\dagger }$, $c_{1}$ or $c_{1}^{\dagger }$ and, finally, $c_{2}$ or $%
c_{2}^{\dagger }$. In particular the equations in (\ref{31}) are already
written in this normal-ordered form. Needless to say, this is an arbitrary
choice and needs not to be, in principle, the \emph{best one}. Here we just
want to remind that normal ordering procedures are rather common in quantum
mechanics for systems with infinite degrees of freedom, and that they have
proved to be quite often useful and reasonable, producing results which are
in good agreement with experimental data.

After some lengthy but straightforward computations we get the following
results:
\begin{equation}
\left\{
\begin{array}{ll}
\sigma _{1}^{(2)}(t)=s_{1}-i\lambda \left( -i\eta
_{1}(t)X_{1}+I_{1}^{s}(t)\right) -i\lambda ^{2}\left( Q_{1}(t)+\overline{%
\eta _{2}(t)}\,Y_{1}\right)  &  \\
\sigma _{2}^{(2)}(t)=s_{2}-i\lambda \left( i\overline{\eta _{1}(t)}%
\,X_{2}+I_{2}^{s}(t)\right) -i\lambda ^{2}\left( Q_{2}(t)+\eta
_{2}(t)Y_{2}\right)  &  \\
\theta _{1}^{(2)}(t)=c_{1}-i\lambda \left( i\overline{\eta _{1}(t)}%
\,X_{3}+I_{1}^{c}(t)\right) -i\lambda ^{2}\left( Q_{3}(t)+\eta
_{2}(t)Y_{3}\right)  &  \\
\theta _{2}^{(2)}(t)=c_{2}-i\lambda \left( -i\eta
_{1}(t)X_{4}+I_{2}^{c}(t)\right) -i\lambda ^{2}\left( Q_{4}(t)+\overline{%
\eta _{2}(t)}\,Y_{4}\right) , &
\end{array}%
\right.   \label{33}
\end{equation}%
where we have proposed the following quantities:
\[
\eta _{1}(t):=\frac{e^{i\hat{\omega}t}-1}{\hat{\omega}},\qquad \eta
_{2}(t)=\int_{0}^{t}\eta _{1}(t_{1})e^{-i\hat{\omega}t_{1}}dt_{1}=\frac{1}{%
\hat{\omega}}\left( t-i\overline{\eta _{1}(t)}\right) ,
\]%
\[
X_{1}:=s_{2}c_{1}c_{2}^{\dagger },\quad X_{2}:=s_{1}c_{1}^{\dagger
}c_{2},\quad X_{3}:=s_{1}s_{2}^{\dagger }c_{2},\quad X_{4}:=s_{1}^{\dagger
}s_{2}c_{1},
\]%
\[
Y_{1}:=s_{1}(c_{1}^{\dagger }c_{1}c_{2}c_{2}^{\dagger }+s_{2}s_{2}^{\dagger
}c_{2}c_{2}^{\dagger }-c_{1}c_{1}^{\dagger }s_{2}s_{2}^{\dagger }),\quad
Y_{2}:=s_{2}(-s_{1}s_{1}^{\dagger }c_{1}^{\dagger }c_{1}+s_{1}s_{1}^{\dagger
}c_{2}^{\dagger }c_{2}-c_{1}c_{1}^{\dagger }c_{2}^{\dagger }c_{2}),
\]%
\[
Y_{3}:=c_{1}(s_{1}s_{1}^{\dagger }c_{2}^{\dagger }c_{2}-s_{2}s_{2}^{\dagger
}c_{2}^{\dagger }c_{2}-s_{1}s_{1}^{\dagger }s_{2}^{\dagger }s_{2}),\quad
Y_{4}:=c_{2}(s_{1}^{\dagger }s_{1}s_{2}s_{2}^{\dagger }+s_{1}^{\dagger
}s_{1}c_{1}^{\dagger }c_{1}-s_{2}^{\dagger }s_{2}c_{1}^{\dagger }c_{1}),
\]%
as well as the following time-dependent operators:
\[
G_{1}(t):=-i\left( -s_{2}c_{1}{I_{2}^{c}(t)}^{\dagger }+s_{2}c_{2}^{\dagger
}I_{1}^{c}(t)+c_{1}c_{2}^{\dagger }I_{2}^{s}(t)\right) ,
\]%
\[
G_{2}(t):=-i\left( s_{1}c_{1}^{\dagger }{I_{2}^{c}(t)}-s_{1}c_{2}{%
I_{1}^{c}(t)}^{\dagger }+c_{1}^{\dagger }c_{2}I_{1}^{s}(t)\right) ,
\]%
\[
G_{3}(t):=-i\left( s_{1}s_{2}^{\dagger }{I_{2}^{c}(t)}+s_{2}^{\dagger }c_{2}{%
I_{1}^{s}(t)}-s_{1}c_{2}{I_{2}^{s}(t)}^{\dagger }\right) ,
\]%
\[
G_{4}(t):=-i\left( s_{1}^{\dagger }s_{2}{I_{1}^{c}(t)}+s_{1}^{\dagger }c_{1}{%
I_{2}^{s}(t)}-s_{2}c_{1}{I_{1}^{s}(t)}^{\dagger }\right) ,
\]%
and
\[
Q_{j}(t):=\left\{
\begin{array}{ll}
\int_{0}^{t}G_{j}(t_{1})e^{i\hat{\omega}t_{1}}dt_{1},\hspace{15mm}j=1,4, &
\\
\int_{0}^{t}G_{j}(t_{1})e^{-i\hat{\omega}t_{1}}dt_{1},\hspace{13mm}j=2,3. &
\end{array}%
\right.
\]%
We can now compute the mean values of ${\sigma _{j}^{(2)}}^{\dagger
}(t)\sigma _{j}^{(2)}(t)$ and ${\theta _{j}^{(2)}}^{\dagger }(t)\theta
_{j}^{(2)}(t)$ on the state $\left\langle \varphi _{\mathcal{G}%
_{0}},\,.\,\varphi _{\mathcal{G}_{0}}\right\rangle $ as seen before. Another
approximation is adopted at this stage: formula (\ref{2a}) shows that the
contribution of the reservoir, $i\gamma _{k}\int_{\mathbb{R}}r_{k}(q)\rho
_{k}(q,t)\,dq$, is $O(\gamma _{k})$ with respect to the other contribution, $%
i_{k}(0)$. For this reason, assuming $\gamma _{k}$ to be small enough, we
approximate $i_{k}(t)$ with $e^{-\left( i\Omega _{k}+\frac{\pi \gamma
_{k}^{2}}{\Omega _{k}^{(r)}}\right) t}i_{k}(0)$.

Up to the second order in $\lambda $, we get
\[
n_{1}(t)\simeq n_{1}+\frac{2\lambda ^{2}}{\hat{\omega}^{2}}(1-\cos (\hat{%
\omega}%
t))[n_{1}(k_{1}n_{2}-k_{1}k_{2}-n_{2}k_{2}-k_{2})+n_{2}k_{1}(1+k_{2})]+%
\lambda ^{2}I_{1}|\eta _{3}^{s}(t)|^{2},
\]%
\[
n_{2}(t)\simeq n_{2}+\frac{2\lambda ^{2}}{\hat{\omega}^{2}}(1-\cos (\hat{%
\omega}%
t))[n_{2}(n_{1}k_{2}-k_{1}k_{2}-k_{1}n_{1}-k_{1})+n_{1}k_{2}(1+k_{1})]+%
\lambda ^{2}I_{2}|\eta _{4}^{s}(t)|^{2},
\]%
\[
k_{1}(t)\simeq k_{1}+\frac{2\lambda ^{2}}{\hat{\omega}^{2}}(1-\cos (\hat{%
\omega}%
t))[k_{1}(n_{1}k_{2}-n_{1}n_{2}-n_{2}k_{2}-n_{2})+n_{1}k_{2}(1+n_{2})]+%
\lambda ^{2}I_{1}|\eta _{3}^{c}(t)|^{2},
\]%
\[
k_{2}(t)\simeq k_{2}+\frac{2\lambda ^{2}}{\hat{\omega}^{2}}(1-\cos (\hat{%
\omega}%
t))[k_{2}(k_{1}n_{2}-n_{1}n_{2}-n_{1}k_{1}-n_{1})+k_{1}n_{2}(1+n_{1})]+%
\lambda ^{2}I_{2}|\eta _{4}^{c}(t)|^{2}.
\]%
Incidentally, these results confirm that the first non trivial contribution
in our perturbation scheme is quadratic in $\lambda $. The following
quantity has been proposed:
\[
\eta _{k}^{s}(t)=\int_{0}^{t}e^{\left[ i(\omega _{k-2}^{s}-\Omega _{k-2})-%
\frac{\pi \gamma _{k-2}^{2}}{\Omega _{k-2}^{(r)}}\right] t_{1}}\,dt_{1}=%
\frac{e^{\left[ i(\omega _{k-2}^{s}-\Omega _{k-2})-\frac{\pi \gamma
_{k-2}^{2}}{\Omega _{k-2}^{(r)}}\right] t}-1}{i(\omega _{k-2}^{s}-\Omega
_{k-2})-\frac{\pi \gamma _{k-2}^{2}}{\Omega _{k-2}^{(r)}}},
\]%
for $k=3,4$. The other function $\eta _{k}^{c}(t)$, is defined like $\eta
_{k}^{s}(t)$ with the only difference that $\omega _{k-2}^{s}$ is replaced
by $\omega _{k-2}^{c}$. If we now compute the variation of the portfolios, $%
\delta \Pi _{j}(t):=\Pi _{j}(t)-\Pi _{j}(0)$, we find that
\begin{equation}
\delta \Pi _{1}(t)=\lambda ^{2}I_{1}\left( |\eta _{3}^{s}(t)|^{2}+|\eta
_{3}^{c}(t)|^{2}\right) ,\quad \delta \Pi _{2}(t)=\lambda ^{2}I_{2}\left(
|\eta _{4}^{s}(t)|^{2}+|\eta _{4}^{c}(t)|^{2}\right) .  \label{34}
\end{equation}%
These formulas show, first of all, that up to the order $\lambda ^{2}$, what
is really important in the computation of the portfolios of the traders, is
not the initial conditions on the cash and shares but, much more than this,
the initial values of the LoI for each trader\footnote{%
If we consider \cite{hav3}, the level of LoI can depend i) on how large the
domain of prices of the payoff function is; ii) the type of payoff function
and iii) the level of public information.}. This is the only \emph{quantum
number} which appears in (\ref{34}), while all the other numbers, $n_{1}$, $%
n_{2}$, $k_{1}$ and $k_{2}$, produce contributions which sum up to zero, at
least at this order in $\lambda $.

Another interesting feature of the analytical expressions for $\delta \Pi
_{j}(t)$ can be deduced observing that,
\[
|\eta _{3}^{s}(t)|^{2}=\frac{e^{-\frac{2\pi \gamma _{1}^{2}}{\Omega
_{1}^{(r)}}t}-2e^{-\frac{\pi \gamma _{1}^{2}}{\Omega _{1}^{(r)}}t}\cos
(\omega _{1}^{s}-\Omega _{1})t+1}{(\omega _{1}^{s}-\Omega _{1})^{2}+\frac{%
\pi ^{2}\gamma _{1}^{4}}{{\Omega _{1}^{(r)}}}^{2}}.
\]%
This implies that $\delta \Pi _{1}(t)$ and $\delta \Pi _{2}(t)$ both admit a
non trivial asymptotic value: calling $\delta \Pi _{j}(\infty
)=\lim_{t,\infty }\delta \Pi _{j}(t)$, and using the above formula for $%
|\eta _{3}^{s}(t)|^{2}$ and the analogous formulas for $|\eta
_{3}^{c}(t)|^{2}$, $|\eta _{4}^{s}(t)|^{2}$ and $|\eta _{4}^{c}(t)|^{2}$, we
get
\begin{equation}
\delta \Pi _{1}(\infty )=\lambda ^{2}I_{1}\left( \frac{1}{(\omega
_{1}^{s}-\Omega _{1})^{2}+\frac{\pi ^{2}\gamma _{1}^{4}}{{\Omega _{1}^{(r)}}%
^{2}}}+\frac{1}{(\omega _{1}^{c}-\Omega _{1})^{2}+\frac{\pi ^{2}\gamma
_{1}^{4}}{{\Omega _{1}^{(r)}}^{2}}}\right) ,  \label{35}
\end{equation}%
and
\begin{equation}
\delta \Pi _{2}(\infty )=\lambda ^{2}I_{2}\left( \frac{1}{(\omega
_{2}^{s}-\Omega _{2})^{2}+\frac{\pi ^{2}\gamma _{2}^{4}}{{\Omega _{2}^{(r)}}%
^{2}}}+\frac{1}{(\omega _{2}^{c}-\Omega _{2})^{2}+\frac{\pi ^{2}\gamma
_{2}^{4}}{{\Omega _{2}^{(r)}}^{2}}}\right) ,  \label{36}
\end{equation}%
The first evident conclusion is that $\delta \Pi _{1}(\infty )+\delta \Pi
_{2}(\infty )\neq 0$. This is possible, since the total amount of cash and
the total number of shares are not required to be constant in time, in our
model. Therefore, there is no reason to expect that the gain for $\tau _{1}$
become the loss for $\tau _{2}$, or viceversa.

As we can see, in agreement with our general analysis in \cite{bagbook}, the
parameters of the free Hamiltonian behave as a sort of inertia for the
system. More in details, if $\omega _{1}^{s}$ and $\omega _{1}^{c}$ are very
large, compared with $\Omega _{1}$ and $\Omega _{2}$, we see that $\delta
\Pi _{1}(\infty )$ is very small: $\tau _{1}$ experiences a large inertia,
so that the value of his portfolio stays almost constant. A similar
conclusion is deduced if $\frac{\gamma _{1}^{2}}{\Omega _{1}^{(r)}}$ is
large enough. Let us now suppose that $\omega _{1}^{s}=\omega
_{1}^{c}=\Omega _{1}$. Then $\delta \Pi _{1}(\infty )=2\lambda ^{2}I_{1}%
\frac{{\Omega _{1}^{(r)}}^{2}}{\pi ^{2}\gamma _{1}^{4}}$. We see from this
formula that the reservoir of the information plays also a role in the
evolution of the portfolios, and we see that, what is relevant for us, is
not really the contribution of the free Hamiltonian, $\Omega _{1}^{(r)}$, or
the contribution of the interaction between the reservoir and the dynamical
variables of the LoI, $\gamma _{1}$, but the ratio above between the two.
This is interesting because it shows that we do have a contribution to $%
\delta \Pi _{j}(\infty )$ coming from these parts of the full Hamiltonian,
even under all the approximations we have considered along the way.

On the other hand, for $\delta\Pi_1(\infty)$ to be large, it is convenient
to have large $I_1$ and/or small values of $\omega_1^s-\Omega_1$, $%
\omega_1^c-\Omega_1$ and of $\frac{\gamma_1^2}{\Omega_1^{(r)}}$. Similar
conclusions can be deduced for $\delta\Pi_2(\infty)$.

A natural question is the following: when does it happen that $\delta \Pi
_{1}(\infty )>\delta \Pi _{2}(\infty )$? This is ensured, for sure, if all
the following inequalities are satisfied:
\[
I_{1}>I_{2},\quad \omega _{1}^{s}-\Omega _{1}<\omega _{2}^{s}-\Omega
_{2},\quad \omega _{1}^{c}-\Omega _{1}<\omega _{2}^{c}-\Omega _{2},\quad
\frac{\gamma _{1}^{2}}{\Omega _{1}^{(r)}}<\frac{\gamma _{2}^{2}}{\Omega
_{2}^{(r)}}.
\]%
Particularly interesting is what happens if $\omega _{1}^{s}-\Omega
_{1}=\omega _{2}^{s}-\Omega _{2}$ and $\omega _{1}^{c}-\Omega _{1}=\omega
_{2}^{c}-\Omega _{2}$. In this case, in order to have $\delta \Pi
_{1}(\infty )>\delta \Pi _{2}(\infty )$, we need to compare two ingredients
of the formulas, i.e. $I_{j}$ and the ratio $\frac{\gamma _{j}^{2}}{\Omega
_{j}^{(r)}}$. As we have seen before, in these conditions $\delta \Pi
_{1}(\infty )>\delta \Pi _{2}(\infty )$ surely if $I_{1}>I_{2}$ and if $%
\frac{\gamma _{1}^{2}}{\Omega _{1}^{(r)}}<\frac{\gamma _{2}^{2}}{\Omega
_{2}^{(r)}}$. But the first inequality implies that the LoI of $\tau _{1}$
should be larger than that of $\tau _{2}$, while the second inequality can
be rewritten as $\frac{\Omega _{2}^{(r)}}{\gamma _{2}^{2}}<\frac{\Omega
_{1}^{(r)}}{\gamma _{1}^{2}}$. This suggests to divide the information
reaching the traders in two different kinds: a \emph{bad information}, which
is directly related to the variables $i_{j}$, $i_{j}^{\dagger }$ and $\hat{I}%
_{j}$, and a \emph{good one}\footnote{%
If we consider \cite{hav3}, the total energy, if it is the harbinger of
public information (relative thus to the payoff function), it will not
necessarily be classified as bad or good information for the portfolio
holder. The diminishing of public information may affect the level of
private information in a different way, if the domain of the payoff function
is small, as opposed to the case when the domain of the payoff function is
large.}, which is related to the reservoir and, therefore, to the variables $%
r_{j}(q)$, $r_{j}^{\dagger }(q)$ and $\hat{R}_{j}(q)$. This is an
interesting result, since it helps to clarify the roles of the different
ingredients of the Hamiltonian (\ref{21}). The differentiation of
information into `good' and `bad' information can also be found back in
early work in finance. The so called `Kyle measure' \cite{Kyle} was proposed
to give an indication of how the level of private information compares to
the level of so called noise trading (which itself is based on a type of
information which is different from private information).

\section{Conclusions}

In this paper we have discussed, within an operatorial setting, a simple
stock market formed by just two traders who, whilst they are interacting
between them, are subjected to a flux of information which aids them to
decide how to behave during the trade operations. A non perturbative result
shows that, in order to not get trivial dynamics, we need to put information
in the model. Otherwise the portfolios of the traders do not change in time.
Using a perturbation expansion we have also deduced the time evolution of
the portfolios of the two traders and we have analyzed their asymptotic
limits at a second order in perturbation theory. This analysis suggests to
contemplate a difference between a bad and a good information. We believe
that this is quite a natural distinction, and it clarifies the meaning of
the various terms in $H$. Interestingly enough, the bad information is
related to a set of two-modes bosonic operators, while the good information
arises from two reservoirs, each having an infinite number of modes.

Needless to say, a step toward real models would imply the following
improvements: more traders, different kind of shares and non constant prices
of the shares. Although the first two extensions do not look particularly
difficult, the last one is very complicated. We hope to be able to produce
such a model in the near future.

\section*{Acknowledgements}

F.B. acknowledges partial financial support from Universit\`{a} di Palermo.
F.B. also wishes to thank the School of Management and Institute of Finance
of the University of Leicester for its warm hospitality.

\end{document}